\documentclass[10pt,conference]{IEEEtran}
\IEEEoverridecommandlockouts
\pdfoutput=1
\usepackage{booktabs}
\usepackage[pdftex]{graphicx}
\usepackage{amsmath}
\usepackage{enumitem}
\usepackage{setspace}
\usepackage{array}
\usepackage{etoolbox}
\usepackage{subfig}
\usepackage{nth}
\usepackage[T1]{fontenc}

\makeatletter
\patchcmd{\@makecaption}
{\scshape}
{}
{}
{}
\makeatletter
\patchcmd{\@makecaption}
{\\}
{.\ }
{}
{}
\makeatother

\usepackage[margin=0.60in]{geometry}

\usepackage{cite}

\usepackage{url}

\makeatletter
\IEEEtriggercmd{\reset@font\normalfont\fontsize{6.8pt}{8pt}\selectfont}
\makeatother
\IEEEtriggeratref{1}

\makeatletter
\let\old@ps@headings\ps@headings
\let\old@ps@IEEEtitlepagestyle\ps@IEEEtitlepagestyle
\def\confheader#1{%
	\def\ps@headings{%
		\old@ps@headings%
		\def\@oddhead{\strut\hfill#1\hfill\strut}%
		\def\@evenhead{\strut\hfill#1\hfill\strut}%
	}%
	\def\ps@IEEEtitlepagestyle{%
		\old@ps@IEEEtitlepagestyle%
		\def\@oddhead{\strut\hfill#1\hfill\strut}%
		\def\@evenhead{\strut\hfill#1\hfill\strut}%
	}%
	\ps@headings%
}
\makeatother

\begin{document}
	\title{\vspace{-15pt}Dynamic Fault Tolerance Through Resource Pooling
	\thanks{Published in the Proceedings of the \nth{13} NASA/ESA Conference on Adaptive Hardware and Systems, UK, 2018. 978-1-5386-7753-7/18/\$31.00 \textcopyright2018 IEEE}}

	\author{\IEEEauthorblockN{Christian M. Fuchs\IEEEauthorrefmark{1},
	Nadia M. Murillo\IEEEauthorrefmark{2},
	Aske Plaat\IEEEauthorrefmark{1},
	Erik van der Kouwe\IEEEauthorrefmark{1},
	and Todor P. Stefanov\IEEEauthorrefmark{1}}
	\IEEEauthorblockA{\IEEEauthorrefmark{1}Leiden Institute for Advanced Computer Science
	\IEEEauthorrefmark{2}Leiden Observatory; \\Leiden University, The Netherlands \hspace*{10pt} email: christian.fuchs@dependable.space\\}
	\vspace{-30pt}}
	
	\maketitle
	
	\begin{abstract}
		Miniaturized satellites are currently not considered suitable for critical, high-priority, and complex multi-phased missions, due to their low reliability.
		As hardware-side fault tolerance (FT) solutions designed for larger spacecraft can not be adopted aboard very small satellites due to budget, energy, and size constraints, we developed a hybrid FT-approach based upon only COTS components, commodity processor cores, library IP, and standard software.
		This approach facilitates fault detection, isolation, and recovery in software, and utilizes fault-coverage techniques across the embedded stack within a multiprocessor system-on-chip (MPSoC).
		This allows our FPGA-based proof-of-concept implementation to deliver strong fault-coverage even for missions with a long duration, but also to adapt to varying performance requirements during the mission.
		The operator of a spacecraft utilizing this approach can define performance profiles, which allow an on-board computer (OBC) to trade between processing capacity, fault coverage, and energy consumption using simple heuristics.
		The software-side FT approach developed also offers advantages if deployed aboard larger spacecraft through spare resource pooling, enabling an OBC to more efficiently handle permanent faults.
		This FT approach in part mimics a critical biological system's ability to tolerate faults, adapt to permanent failure, and enables graceful aging of an MPSoC.
	\end{abstract}
	
	\section{Introduction}
	Satellite miniaturization has enabled a broad variety of scientific and commercial space missions, which previously were technically infeasible, impractical or simply uneconomical.
	However, very small satellites such as nanosatellites and sometimes even microsatellites ($\leq$100kg) are currently not considered suitable for critical and complex multi-phased missions, as well as high-priority science applications, due to their low reliability.
	On-board computer (OBC) and related electronics constitute a large part of such a spacecraft's mass, yet these components lack often even basic fault tolerance (FT) functionality.
	Due to budget, energy, mass and volume restrictions, existing FT solutions originally developed for larger spacecraft can in general not be adopted.
	Nanosatellite OBCs also have to cope with drastically varying workload throughout a mission, which traditional FT solutions can not handle efficiently.
	Therefore, we developed a novel FT approach offering strong fault coverage, which was implemented fully using only a single FPGA with commodity processor designs, and library IP. 
	
	This architecture can protect generic applications with an arbitrary structure, can adapt to varying performance requirements in longer multi-phased missions, and can adapt to a shrinking pool of processing capacity similar to a biological system, efficiently handling aging effects and accumulating permanent faults.
	As major parts of our approach are implemented in or directly controlled by software, a spacecraft operator can configure the OBC to deliver the desired combination of performance, robustness, functionality, or to meet a specific power budget.
	To offer strong fault detection, isolation and recovery (FDIR), we combine software-side fault detection and mitigation and configuration scrubbing with various other FT measures across the embedded stack, enabling strong, low-cost FT with commodity hardware, while exploiting FPGA reconfiguration to mitigate permanent faults.
	
	The next two sections contain background information, and a discussion of related work. 
	In Section \ref{sec:introandterminology} a brief overview over the three stages of our approach is provided.
	Our proof-of-concept OBC-design is described in Section \ref{sec:mpsoc}, with the functionality of each FT-stage outlined in the subsequent sections.
	How this approach can improve efficiency of OBC in spacecraft of all weight classes, spare resource utilization and fault coverage, is discussed in Section \ref{sec:FaultHandlingAndSpares}.
	Section \ref{sec:remapping}, introduces \emph{performance profiles} allowing a system-on-chips (SoC) to trade compute performance for energy efficiency, robustness, and functionality at runtime.
	Our approach provides advantages to spacecraft of all weight classes, and can be implemented also within distributed systems, for which further applications and improvements are discussed in Section \ref{sec:discussion}.
	
	\subsection*{Contributions:}
	\begin{itemize}[leftmargin=10pt]
		\item An architecture enabling software-side fault detection and mitigation for commercial-off-the-shelf (COTS) MPSoCs and FPGA-based systems, implementing the full FDIR cycle and combining FT measures across the embedded stack.
		\item A practical solution to safely repurpose redundancies in an OBC by exploiting mixed criticality, allowing spare resource pooling, thereby increasing fault coverage capacity and reducing the need for over-provisioning.
		\item Functionality allowing an OBC to deliver a runtime configurable level of performance by dynamically trading fault coverage, processing capacity, and energy consumption.
	\end{itemize}
	
	\section{Background}
	Tasks which would be handled by multiple dedicated payload and subsystem processing systems aboard a larger satellite, are usually handled by just one COTS-based command \& data handling system in nanosatellites.
	These utilize mobile-market and embedded SoCs with one or more cores (MPSoCs), SDSoCs \cite{carlson2016use}, or FPGAs \cite{kastensmidt2016fpgas}.
	Due to manufacturing in fine technology nodes, such chips offer superior efficiency and performance as compared to space-grade OBC designs, but are also non-FT\footnote{Exceptions to this rule received uncommonly abundant funding, are technology demonstration for FT concepts, or custom fail-over designs.}. 
	These SoCs consist mostly of extensively tested and optimized standard logic, reused, supported, and evolved continuously by several industries and used daily by countless developers.
	In contrast, most radiation-hard-by-design (RHBD) processors cores, and SoCs manufactured in more robust manufacturing processed (RHBM) are crafted almost artisanally at high cost by few designers with little commercial stimulus for optimization.
	Their cost, energy consumption and mass often exceed such a spacecraft's global power budget, total mass, and almost always its overall project budget.
	Therefore, we developed a hybrid FT-approach based upon only COTS components, library IP, and existing software, instead of artisanal processor designs and proprietary instruction set architectures.
	
	Existing hardware voting based FT solutions are design-time static and can tolerate a fixed number of failures within a voter setup, which can not be changed at runtime.
	Critical biological systems instead consist of independent, cooperating cells or clusters of similar functionality with a high degree of inherent redundancy and self-healing capabilities.
	Damage to a single cell is compensated by the remaining cells, and a complete breakdown of functionality occurs only due severe damage to the system at a broader scale.
	Our approach combines various FT techniques to mimic such behavior at the logic and SoC level, through FPGA reconfiguration and software-controlled thread migration within a globally share pool of processor cores, enabling graceful aging.
	The replication level, hence fault coverage capabilities, and various other parameters can be adjusted at runtime, while spare capacity can be reused to run background and lower-criticality applications instead of remaining idle.
	
	In low feature-size chips, the energy threshold above which highly charged particles can induce faults in digital logic (single event effects - SEE) decreases, while the ratio of events inducing multi-bit upsets (MBU), and the likelihood of permanent faults in logic and memory increases.
	Increased fault coverage of hardware-FT based concepts on such chips through additional FT-circuitry therefore implies diminishing returns, preventing an application of traditional RHBD/RHBM concepts \cite{reick2008fault, hijorth2015gr740} to mobile-market SoCs.
	Total ionizing dose, however, becomes less of a problem with finer technology nodes, and recent generation FPGAs also show decent latch-up performance \cite{berg2015nasaFPGAtestOverview, Tambara2015zynqGoodRadTestPerformance}.
	FPGAs have drastically improved FDIR potential \cite{Wirthlin2015HighReliabilityFPGA} despite being more vulnerable to transients, as radiation-induced upsets in the running configuration can be corrected via reconfiguration with differently routed configuration variants \cite{bozzoli2017self}.
	
	\section{Related Work}
	\label{sec:relatedwork}
	Fine-grained, non-invasive, and scalable fault detection in FPGA fabric is challenging, and subject of ongoing research \cite{ebrahimi2016low, rittner2017automated}, and often is simply ignored \cite{martinez2017fully}.
	Most FPGA-based FT-concepts rely on error scrubbing, which has scalability limitations for complex logic \cite{ebrahimi2016low, stoddard2017hybrid}, unless special-purpose offline testing is utilized \cite{siegle2016availability}.
	In the future, memory-based reconfigurable logic devices (MRLDs) \cite{wang2017testingMRLD} may allow programmed logic to be protected like conventional memory, and thus would drastically simplify fault detection.
	If manufactured using phase/polarity-change memory instead of charge-based technologies, MRLDs could further increase robustness, but this technology is only today being productized.
	In this paper, we thus present an approach to general-purpose FT computing that compensates for faults across the embedded stack and through partial FPGA reconfiguration.
	We realize fine-grained fault detection at the software level, and perform scrubbing only as an auxiliary measure in the background to increase robustness of our SRAM-based FPGA platform.
	
	Hardware voting today is used exclusively for protecting simpler FT processor cores at the microcontroller level \cite{hijorth2015gr740, iturbe2016triple}, and for accelerators \cite{guerrieri2018sefuw} supporting application code with tightly constrained program structure.
	Hence, the application of this hardware-centered approach has become a technical dead-end for protecting widely used application processor designs intended for general-purpose computing, while accelerators by themselves would only assure FT for computation and data offloaded to such a device.
	In our research, however, we seek to deliver strong fault coverage for general purpose computing, and aim to efficiently protect even larger and more complex modern application processors, such as those widely used in mobile market and embedded devices.
	Mobile market processors can run at gigahertz clock rates, for which hardware-side voting or instruction-level lockstep are non-trivial, hence, hardware voting approaches have been implemented only at lower clock rates \cite{iturbe2016triple, decoursey2006non, pigno2011testbench}.
	For comparison, today's highly optimized COTS library IP achieves clock speeds comparable to traditional FT-processor designs on ASIC even on an FPGA, without requiring manual fine-tuning.
	We instead utilize software-driven coarse-grain lockstep to achieve fault detection, and maintain consistency between cores, requiring no vast arrays of synchronized voters, while utilizing COTS IP.
	
	Thread migration has been shown to be a powerful tool for assuring FT, but prior research ignores fault detection, and imposed tight constraints on an application's type and structure (e.g., video streaming and image processing \cite{martinez2017fully}).
	However, to implement sophisticated and efficient thread migration, fault-detection must be facilitated at the OS or application-level without falling back to design space exploration.
	Coarse-grain lockstep of weakly coupled cores can do just that, and in the past has already been used for high availability, non-stop service, and error resilience concepts.
	However, in prior research, faults are usually assumed to be isolated, side effect free and local to an individual application thread \cite{holler2015software} or transient \cite{dobel2014operating, munk2015toward}, and entail high performance \cite{santangelo2013open} or resource overhead \cite{missimer2014distributed, al2016fault}.
	More advanced proof-of-concepts \cite{kretzschmar2016synchronization, dobel2014operating}, however, attempt to address these limitations, and even show a modest performance overhead between 3\% and 25\%, but utilize checkpoint \& rollback or restart mechanics \cite{dobel2014operating}, which make them unsuitable for spacecraft command \& control applications.
	
	\section{System Overview \& Requirements}
	\label{sec:introandterminology}
	Coarse-grain lockstep is one among several measures used in our hybrid FT approach to facilitate forward-error-correction (FEC) and deliver strong fault coverage.
	Our approach consists of three fault mitigation stages:
	
	\textbf{Stage 1} utilizes coarse-grain lockstep for fault detection, to generate a distributed majority decision between processor cores.
	Stage 1 utilizes time-triggered checkpoints to autonomously resolved faults corrupting the state of applications, and facilitate re-synchronization and thread migration in case of repeated faults, enabling strong \textbf{short-term fault coverage}.
	
	\textbf{Stage 2} assures the integrity of programmed logic by interfacing with Stage 1 and functionality such as Xilinx SEM.
	Its objective is to assure and recover the integrity of processor cores and their immediate peripheral IP through FPGA reconfiguration, thereby \textbf{counteracting resource exhaustion}.
	
	\textbf{Stage 3} handles resource exhaustion and re-allocates processing time within the system to \textbf{maintain stability of critical applications and functionality in a degraded system}.

\begin{figure}[!t]
		\centering
		\includegraphics[width=1\linewidth]{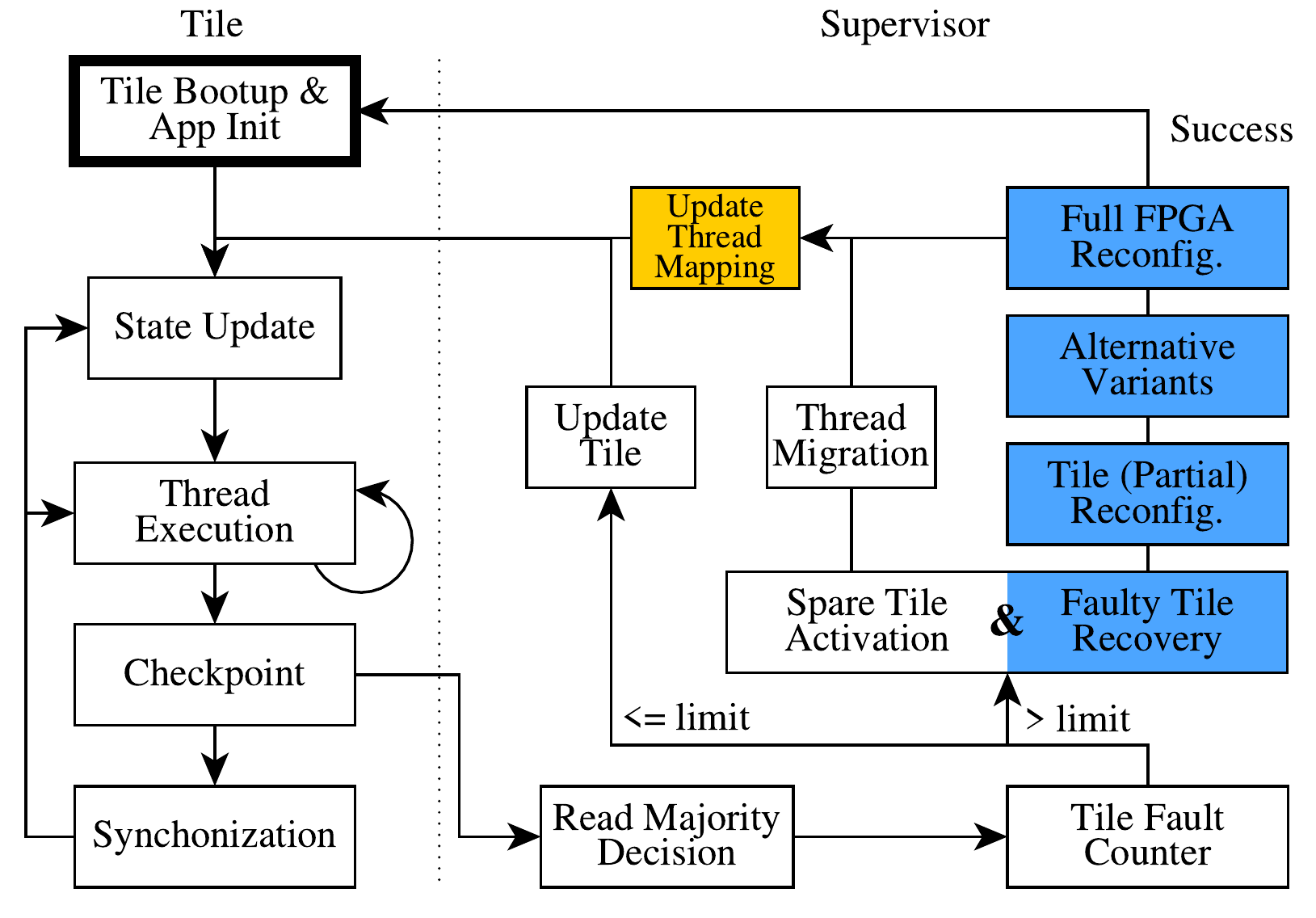}
		\vspace{-10pt}
		\caption{Stage 1 (white) implements a continuous checking loop, which facilitates fault coverage through thread-level synchronization and migration between tiles.
			Stage 2 (grey) can recover faulty tiles using reconfiguration.
			In case of resource exhaustion, Stage 3 adapts the thread allocation to best utilize the remaining processing capacity.}
		\label{fig:supervision}
		\vspace{-5pt}
	\end{figure}

	The entire Stage1-3 form a closed cycle, which implements FDIR in several steps as depicted in Figure \ref{fig:supervision}.
	Additional implementation details on Stage 1's thread-level coarse-grain lockstep, beyond what is briefly described in Section \ref{sec:stage1} is available in \cite{fuchs2017atATS}.
	
	In low-end nanosatellites (e.g. 1U CubeSats), Stages 1$+$3 can be implemented separately on a generic MPSoC, providing a level of system-level robustness which otherwise would be only be achievable through proprietary hardware-FT solutions.
	For all other spacecraft, we complement this functionality with a tiled MPSoC architecture for FPGA as outlined in the next section, which allows the system to recover defective tiles through reconfiguration, and enables it to more efficiently handle permanent faults.
	
	\section{A Dynamic Tiled MPSoC Architecture}
	\label{sec:mpsoc}
\begin{figure}[!t]
		\centering
		\includegraphics[width=0.8\linewidth,clip, trim = -10pt 0pt 10pt 0pt]{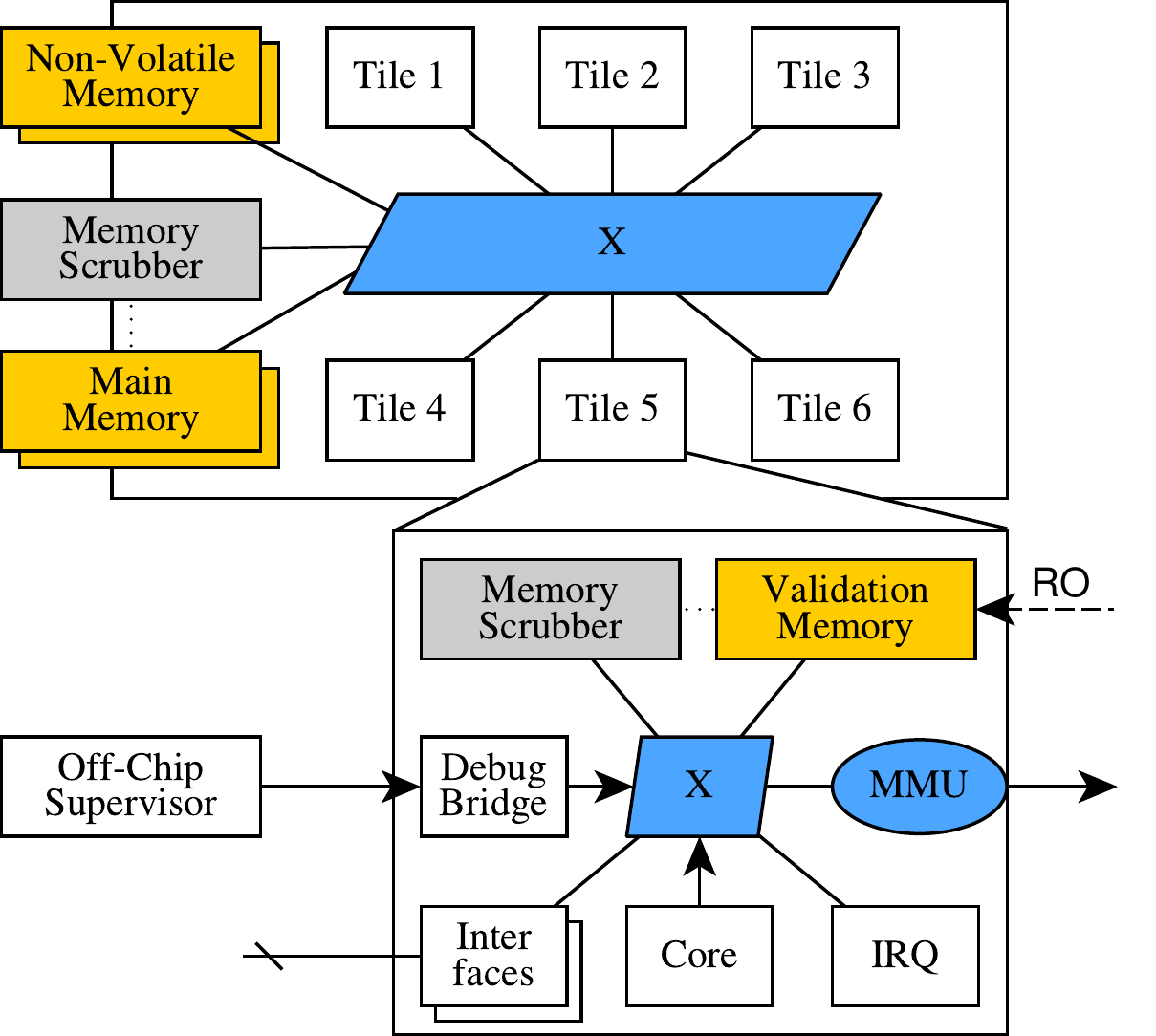}
		\caption{A high-level topology diagram of our tiled MPSoC architecture with memory controllers highlighted in yellow, and interconnect-logic in blue. A debug-bridge on each tile allows supervisor access. Access to each tile's validation memory is possible read-only through the global interconnect.}
		\label{fig:mpsoc}
		\vspace{-10pt}
	\end{figure}

	Figure \ref{fig:mpsoc} depicts a simplified and publicly reproducible version of our MPSoC design.
	It follows a tiled architecture with each tile containing a processor core, local interconnect, and peripheral IP-cores and interfaces.
	A debug bridge allows supervisor access to each tile, e.g., to perform introspection for testing purposes or to trigger a reset.
	The only globally shared resources are a set of redundant main memory controllers and non-volatile (nv) data storage.
	Code in nv-memory can be shared between tiles, while widely used DDR and SDRAM controllers are too large to instantiate for each tile, and would require an excessive number of I/O-pins.
	Hence, our MPSoC architecture consists of isolated SoC-compartments accessing shared main memory and operating system code, in contrast to the conventional MPSoC designs, where cores share most infrastructure and peripherals.
	
	Our main platform is the commercial ARM Cortex-A53 application processor core, which was chosen due to its flexibility, wide-spread use in mobile-market MPSoCs and scalability.
	The Xilinx Zynq Ultrascale+ SDSoC, also contains four discrete A53 cores and is foreseen to launch in one of our main target missions.
	The design outlined in this paper was facilitated using Xilinx/Microblaze IP, and Microblaze-specific IP is replaced with ARM equivalents.
	These are freely available as part of the Xilinx IP-Library, and due to its maturity, flexibility, broad OS support (for Linux, RTEMS, FreeRTOS), and are widely availability.
	Hence both Microblaze and Cortex-A53 cores are both solid choices for low-cost nanosatellite applications, with a Cortex-A53 cores offering better absolute performance.
	
	Each tile's checkpoint-related information is stored in a dedicated on-chip dual-port BRAM memory (\emph{validation memory}) and exposed to other tiles, to allow low-latency information exchange between tiles without requiring inter-tile cache-coherence or access to main memory.
	Validation memory is writable through the tile-local interconnect, and is read-only accessible by other tiles.
	
	The address space layout on each tile, including mapping of tile-private peripherals and interfaces are identical.
	Each tile can access its own main memory address segment, which is mapped to the same address range on all tiles.
	Additionally, main memory in its entirety (all memory segments) is read-only accessible system wide, to simplify state synchronization between tiles.
	
	All tiles are equipped with the same interface configuration, with controllers being mapped to identical locations in address ranges.
	Hence, the address space layout is uniform across all tiles in the system.
	Therefore, application code and data structures are portable between tiles, simplifying thread migration drastically, and allowing direct re-use of many data structures.
	Full replication of all interfaces across all tiles is not required, but simplifies thread assignment and development manpower.
	Tiles can be made aware of varying interface configurations per tile to reduce the MPSoC's footprint, fault potential, and I/O pin count, but this is beyond the scope of this paper.
	
	Caches, on-chip BRAM, and globally-shared memories are ECC protected.
	Xilinx Library IP already offers SECDED coding for on-chip BRAM and Microblaze caches, whereas most recent Cortex cores foresee stronger ECC for caches.
	Main memory is conventional DDRx-SDRAM with ECC, whereas radiation-tolerant FeRAM \cite{zhang2015single} is used as nv-memory for operating system data and code, and COTS MLC-NAND-Flash for data storage.

	For nanosatellite missions to LEO, a variety of DDR memory controllers with ECC support are available as part of standard vendor IP-libraries.
	For more deep-space and long-term missions, stronger erasure coding should be used due to the increased impact of SEEs and higher likelihood of MBU in high-density SDRAM.
	Relevant well tested and proven controllers implementing Reed-Solomon block coding are available commercially, or can be assembled from generic ECC-IP and standard controller cores\footnote{all necessary cores are available open-source e.g. from OpenCores}.

	ECC-error syndromes generated within a tile, are handled locally.
	Syndromes in validation memory generated due to access by other tiles during a checkpoint are deferred and processed after the checkpoint.
	Syndromes from globally shared controllers can either be handled by the supervisor, or passed through to all tiles and masked based on the related address; this is a design decision.
	For simplicity, ECC syndromes in main memory are passed on to the supervisor in our proof-of-concept.
	
	We implemented this MPSoC successfully on current generation Xilinx Zynq/Kintex and Virtex FPGAs with 4, 6 and 8 tiles.
	Tiles can be placed on separate configuration partitions to enable partial reconfiguration of individual tiles, without affecting the rest of the system.
	A positive side effect of such floor planning is a strong spatial separation of tile logic, thereby reducing the likelihood of MBUs corrupting more than a single tile.
	Further information on this implementation including floor planning, and a detailed utilization report can be found in \cite{fuchs2017performance}.

	In \cite{fuchs2017performance}, we also conducted a series of benchmarks of our lockstep implementation to estimate the performance impact of our approach.
	As the overhead of software-side FT measures has been shown in literature to vary broadly between very low 3\% and extreme 25\%, we intentionally chose extreme parameters for our benchmark application and drastically increased the checkpoint frequency to 20hz.
	For comparison, based on radiation testing data for Xilinx Ultrascale FPGAs, we would today consider checkpoint periods of once per second to once ever 5 seconds reasonable in low-earth orbit.
	Note that we intentionally chose to carry out benchmarking in user-land, not within our RTOS/RTEMS-based implementation running bare-metal.
	Our approach is interrupt and context-switch heavy, and utilizes a considerable amount of thread-management calls.
	In user-land, such operations imply one or multiple system calls in addition to the actually executed function.
	This increases the computational cost of such operations drastically as compared to our actual bare-metal implementation, and was done on purpose to add an additional margin for overhead, to achieve an upper bound of our approach's performance cost.
	
	We deployed erasure coding based configuration error mitigation using Xilinx Soft-Error-Mitigation for Ultrascale FPGAs (SEM) and supervisor-side scrubbing safeguard logic integrity.
	However, SEM and scrubbing only address specific faults in certain parts of an FPGA, and leave large parts of logic unprotected.
	Therefore, the software-side functionality outlined in the next sections closes this protective gap.

	During a checkpoint, the state of a all threads mapped to a tile is compared and synchronized with its siblings.
	To do so, the checkpoint handler executes an application-provided callback function for all pending threads, producing checksums generated from thread-private data structures.
	Checksums are stored in the tile's local validation memory and thereby exposed to the other tiles, and then compared with the other tiles in the system.
	In case of disagreement, the tile signals disagreement with that sibling and executes synchronization callbacks for all affected threads.
	If necessary, it then also executes relevant update callbacks and then resumes application execution.
	We published an in depth description of these mechanics as well as benchmark results for an astronomical application in \cite{fuchs2017atATS}.
	
	\section{Stage 1: Short-Term Fault Mitigation}
	\label{sec:stage1}
	The objective of Stage 1 is to detect and correct faults within a tile, and assure a consistent system state through checkpoint-based FEC.
	It is implemented as sets of tiles running two or more copies of application threads (siblings) in lock step.
	Checkpoints interrupt execution, facilitating the lockstep and enforcing synchronization, allowing thread assignment within the system to be adjusted if required, as depicted in Figure \ref{fig:supervision}.
	
	This approach enables us to utilize application intrinsics to assess the health state of the system without requiring in-depth knowledge about the application code.
	The supervisor just reads out the results of the tiles' decentralized consistency decision.
	Threads can be scheduled and executed in an arbitrary order between two checkpoints, as long as their state is equivalent upon the next checkpoint.

	We avoid thread synchronization issues due to invasive lockstep mechanics \cite{kretzschmar2016synchronization} by merely reusing existing OS functionality without breaking existing ABI contracts.
	Therefore, we can continue relying upon pre-existing synchronization mechanics such as POSIX cancellation points\footnote{e.g. sleep, yield, pause, for further details, see IEEE Std 1003.1-2017 p517} and their bare-metal equivalents (e.g., \emph{RTEMS\_NO\_PREEMPT} in RTEMS's Classic API if used instead of \emph{newlib} or the POSIX API).
	Stage 1 can even deliver real-time guarantees, and the tightness of the RT guarantees depends upon the time required to execute application callbacks.
	In our RTEMS/POSIX-based implementation, we utilize priority-based, preemptive scheduling with timeslicing, allowing threads to delay checkpoints until they reach a viable state for checksum comparison.
	
	Checkpoints are time triggered, but can also be induced by the supervisor through an interrupt e.g. to signal that new threads have been assigned.
	Thus, the OS only has to support interrupts, timers, and a multi-threading capable scheduler.
	To the best of our knowledge, such functionality is available in all widely used RT- and general purpose OS implementations.

	A fault resolved during a checkpoint may cause the affected tile to emit incorrect data through I/O interfaces, an inherent limitation to coarse-grain lockstep \cite{dobel2014operating}.
	For many very small nanosatellite missions this is acceptable, as the use of COTS components requires incorrect I/O to be sanitized anyway.
	In contrast, larger spacecraft already utilize interface replications or even voting, usually requiring considerable effort at the interface level to facilitate this replication.
	Our approach combined with the previously described MPSoC architecture inherently provides interface-level replications by design, no longer requiring extra measures to be taken.
	Additional protection is therefore only needed for space applications where non-propagation of incorrect I/O is required but interface replication is undesirable, i.e., due to PCB-space constraints aboard CubeSats or unchangeable subsystem requirements.
	For packet-based interfaces such as Spacewire, AFDX, CAN, or Ethernet, no hardware-side solution is necessary, as data duplication can be managed more efficiently at OSI layer 2+.
	This approach today is widely used as part of real-time capable FT-networking \cite{AFDX}.
	Other interfaces like I2C and SPI allow a simple majority decision per I/O line, which can be implemented on-chip through FIFO buffers, as the remaining on-tile interfaces have low pin count and run at relatively low clock frequencies.
	
	\section{Stage 2: Tile Repair \& Recovery}
	\label{sec:stage2}
	Stage 1 can not reclaim defective tiles, eventually resulting in resource exhaustion.
	Therefore, in this stage, we recover defective tiles through reconfiguration to counter transients in FPGA fabric.
	To do so, the supervisor will first attempt to recover a tile using partial reconfiguration.
	Afterwards, the supervisor validates the relevant partitions to detect permanent damage to the FPGA (well described in, e.g., \cite{nguyen2017repairing}), and executes self-test functionality on the tile to detect faults in the tile's main memory segment and peripherals.
	If unsuccessful, the supervisor can repeat this procedure with differently routed configuration variants, potentially avoiding or repurposing permanently defective logic.
	
	As tiles are placed along partition borders in our MPSoC architecture, tiles can be recovered in the background without interrupting the rest of the system.
	The supervisor can also attempt full reconfiguration implying a full reboot of all tiles.
	Further details on reconfiguration and error scrubbing with a microcontroller-based proof-of-concept implementation for a nanosatellite are available in \cite{fuchs2016enhancing}.
	If both partial- and full-reconfiguration are unsuccessful and all spare resources have been exhausted, Stage 3 is utilized to assure a stable system core to enable operator intervention.
	
	\section{Stage 3: Applied Mixed Criticality}
	\label{sec:stage3}
	Stage 3 autonomously maintains system stability of an aged or degraded OBC.
	When considering a miniaturized satellite's OBC, we can differentiate individual applications or parts of flight software by criticality.
	At the very least, we will find software essential to a satellite's operation, e.g. platform control and commandeering, as well as other applications of various levels of lower criticality.
	If the previous stages no longer have enough spare processing capacity or tiles to compensate the loss of a tile, this stage utilizes thread-level mixed criticality to assure stability of core OBC functions.
	To do so, it can sacrifice lower criticality tasks in favor of providing compute resources to reach the desired replication level for critical threads.
	
	Dependability for higher-criticality threads efficiently can be maintained by reducing compute performance or reliability of lower-criticality applications.
	Lower-criticality tasks may be executed less frequently or on fewer tiles, thereby reducing functionality or fault coverage for these tasks, retaining resources for higher-criticality threads.
	This decision is taken autonomously, and the operator can then define a more resource conserving satellite operation schedule at a spacecraft level, e.g., sacrifice link capacity, or on-board storage space, to make best use of the OBC in its degraded state.
	
	\section{Spare Resource Pooling}
	\label{sec:FaultHandlingAndSpares}
	This FT approach enables FT even for very small satellites, but provides benefits for spacecraft of all weight classes.
	To increase fault coverage in traditional hardware voting FT systems, additional cores and spares must be provisioned, while compute performance can be increased by utilizing faster processors cores and adding more hardware voting instances.
	This is done at design time, requiring over-provisioning, and can not be changed throughout a mission.
	Cores are hardwired to a specific instance, therefore, an instance will degrade once its spares are exhausted, even if idle spares were available elsewhere.

	In contrast, our approach is not based on hardwired voting instances, as applications are mapped to a global pool of tiles with a given replication level.
	In principle, our approach does utilize spare resources too, but spare tiles do not differ from conventional tiles in any way.
	Hence, spare tiles do not have to remain idle, and unused processor capacity becomes a spare resource that can be re-purposed.
	Thus, the fault coverage capabilities of the system are no longer dependent on the distribution and location of permanent faults within the system, increasing overall robustness.

	As applications can be migrated between tiles, low criticality threads and background tasks can be assigned to utilize free spare capacity.
	These lower-criticality threads can be de-scheduled in favor of higher-criticality applications, if needed.
	Spare capacity can also be used to increase FT for threads, which usually would be executed without majority voting or separately due to resource constraints.
	We can distribute a defective tile's workload to other tiles, to best take advantage of the remaining system resources.
	
	The best target tiles and to-be-evicted threads are not determined ad-hoc, but before a fault actually occurs, to reduce the time spent in a checkpoint.
	We can maintain one replacement strategy for every tile, due to the low tile and thread counts common in space applications today\footnote{Manycore systems would allow too many combinations, but they will not be applied to on-board data handling in the foreseeable future.}.
	Subsequent to a fault, these strategies are recomputed to consider the now reduced processing capacity of the system.
	As thread assignments are not controlled by the supervisor, but only adjusted, threads may exit, fork or create new child threads.
	Therefore, an update to adjust these strategies to the currently running threads is also triggered based on the fault counter mechanics of Stage 2.
	Even if a fault occurs immediately after the current checkpoint, these strategies will only be needed at the next checkpoint.
	Therefore, this is a background operation which can be handled by the supervisor, allowing the OBC to resume processing immediately.

	\begin{figure}[!b]
	\vspace{-20pt}
	\centering
	\subfloat[Migration by low-criticality thread pruning.]{
		\includegraphics[width=0.90\linewidth]{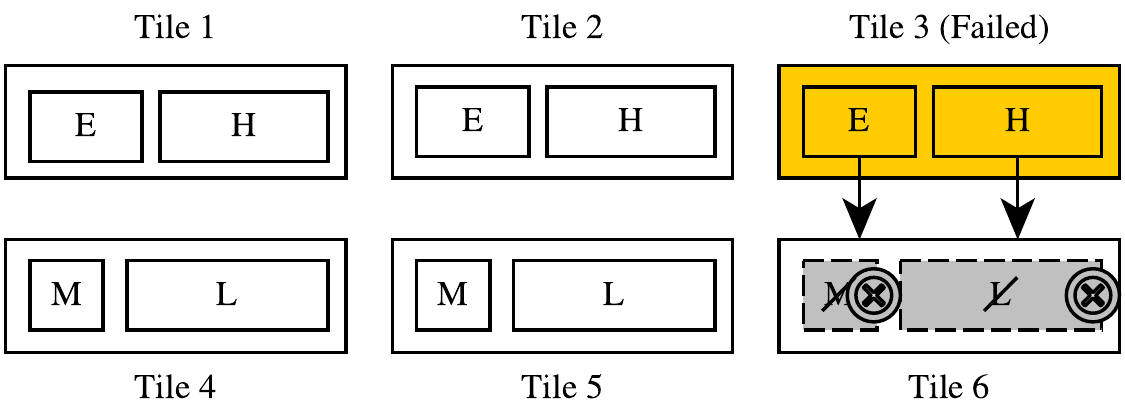}
		\label{fig:6mpsocfaultreplace}
	}%
	\vspace{-5pt}
	\\
	\subfloat[Migration through clock-speed increase.]{
		\includegraphics[width=0.90\linewidth]{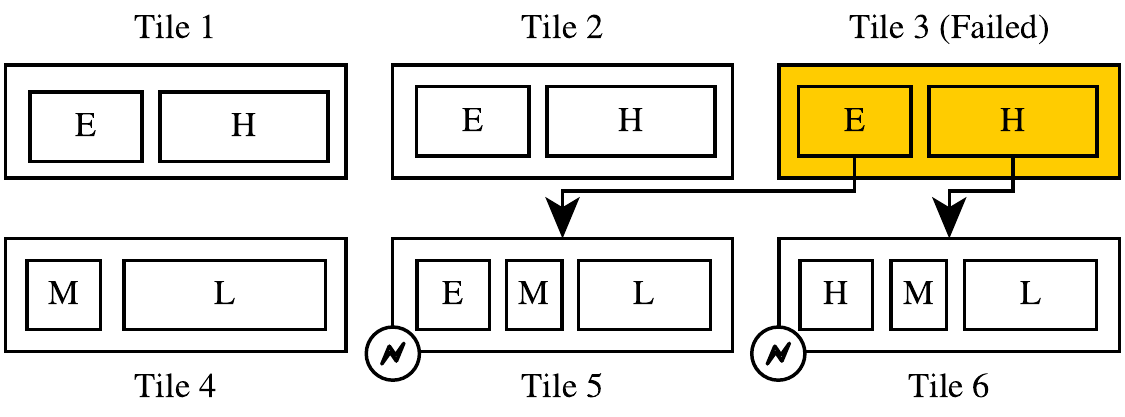}
		\label{fig:6mpsocsplitworkto3and4}
	}%
	\vspace{-5pt}
	\\
	\subfloat[Migration through processing time reduction.]{
		\includegraphics[width=0.90\linewidth]{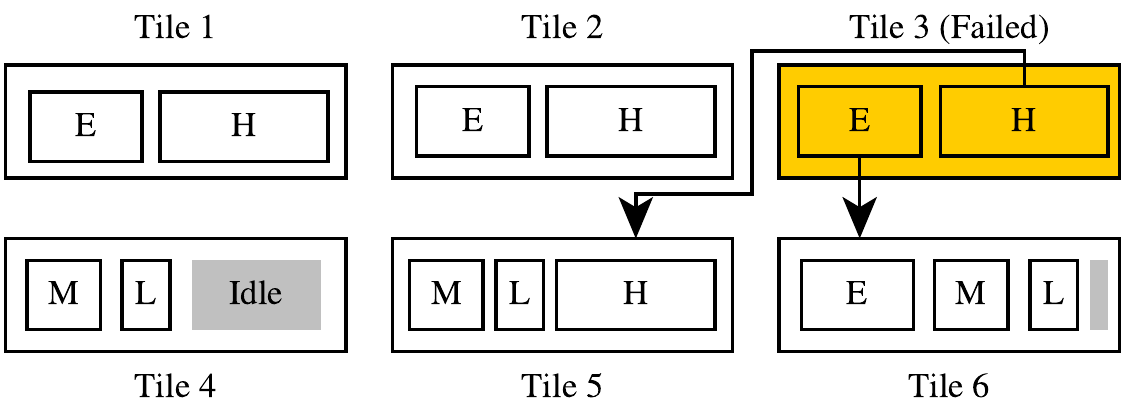}
		\label{fig:6mpsocdecreasetimeforlow}
	}%
	\caption{A 6-tile MPSoC running 4 threads of mixed criticality (\textbf{E}ssential, \textbf{H}igh, \textbf{M}edium, and \textbf{L}ow), where tile 3 (yellow) suffered a hard fault.
	To retain majority voting for the higher criticality threads, different recovery strategies can be facilitated through, without directly requiring spares.}
	\label{fig:6mpsocfigure}
	\vspace{-10pt}
	\end{figure}

	Figure \ref{fig:6mpsocfigure} depicts a six tile MPSoC running four applications of different criticality.
	A fault has occurred in tile 3, which has been marked as permanently defective, and there are multiple recovery solutions:
	\begin{itemize}[leftmargin=10pt]
		\item Affected threads could be relocated to a tile running lower-criticality applications, replacing them as depicted in Figure \ref{fig:6mpsocfaultreplace}.
		For example, the threads previously run on tile 3 can be migrated to tile 6, replacing lower criticality thread-copies previously run there.
		This requires tile 6 to copy the state of its newly assigned threads from tile 1 or 2, at the cost of executing the lower-criticality applications redundantly instead of with majority voting.
		\item Instead of entirely de-scheduling one instance of each lower criticality threads, the clock frequency on two tiles could be increased, allowing one of each high-criticality thread to be migrated.
		In Figure \ref{fig:6mpsocsplitworkto3and4}, this is depicted by moving the threads from the failed tile to tiles 5 and 6 without de-scheduling instances of the low criticality threads.
		This is possible as coarse-grain lockstep only requires an equivalent state between siblings upon reaching a checkpoint and no cycle-accurate synchronization.
		Most modern embedded and mobile-market cores support frequency scaling.
		\item Another possibility would be to instead increase the clock frequency of just one tile, if sufficient additional processing capacity can be made available that way.
		\item Finally, in contrast to increasing the clock frequencies of individual tiles, tile 4-6's schedulers could also assign less processing time to the lower-criticality tasks as shown in Figure \ref{fig:6mpsocdecreasetimeforlow}.
		Due to timing implications for real-time applications, this may only be possible for sporadic tasks, and background applications, which do not require a fixed amount of processing time.
		Also, to guarantee equivalent work is conducted for the medium and lower-criticality threads, the schedulers on 3 instead of just 2 tiles would require adjustment, wasting processing capacity in Tile 4 and 6.
		However, during this idle time, Tile 4 could be deactivated to reduce energy consumption.
	\end{itemize}
	The ideal recovery strategy, depends on the current performance requirements towards the OBC.
	Additional thoughts on this aspect are discussed, e.g., in \cite{martinez2017fully}, where different replacement strategies are described at a more mathematical level for video streaming applications.
	In the next section, we therefore discuss a heuristic approach to find near-best solutions to calculate this decision autonomously and rapidly, considering different performance requirements.

	\section{Adapting to Varying Mission Requirements}
	\label{sec:remapping}
	
\begin{figure}[!t]
		\vspace{-5pt}
		\centering
		\includegraphics[width=1\linewidth,clip, trim = 35pt 0pt 28pt 148pt]{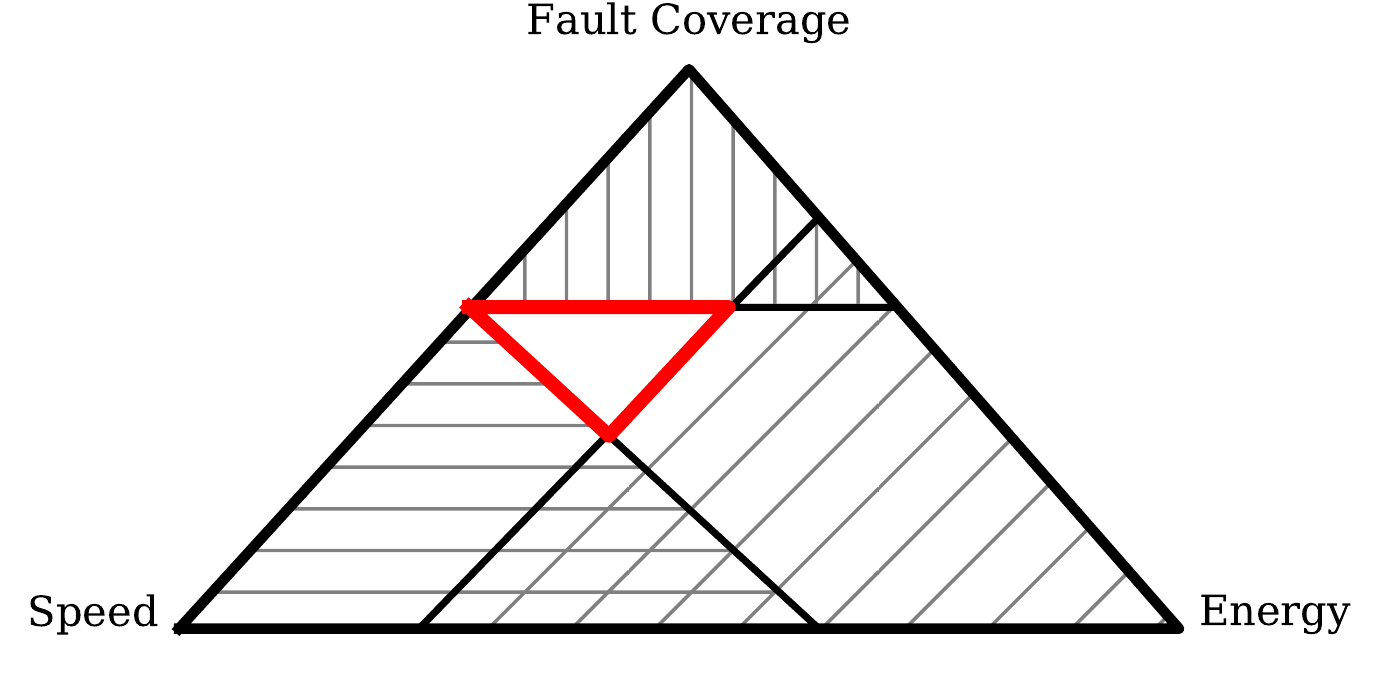}
		\vspace{-20pt}
		\caption{An MPSoC utilizing the presented approach can trade speed, energy efficiency, and fault coverage at run-time.
			We utilize \emph{performance profiles} for each objective to facilitate a heuristic solution, to approximate the ``best possible" set of thread-mapping in the highlighted area.}
		\label{fig:FtSpeedEnergyTriangle}
		\vspace{-15pt}
	\end{figure}

	The approach described in the previous sections allows an OBC to meet a desired power budget, maximize fault coverage, processing power, or even functionality.
	Hence, the spacecraft can better fulfill its scientific or commercial mission, and increase the spacecraft's lifetime.
	Theoretically, all we need to do is find the ideal set of thread mappings which fulfill our desired trade-off between processing capacity, FT, and minimal energy consumption.
	These three performance objectives can be visualized as depicted in Figure \ref{fig:FtSpeedEnergyTriangle}, and viable mappings can be found in the inner area outlined in red.
	
	These three objectives oppose each other, and fully dynamic performance optimization at runtime is non-trivial and costly. 
	Computer science usually approaches such issues with computationally expensive optimization algorithms to find the ideal solution, or design space exploration to find a large set of near-best and chose the optimal solution either at runtime \cite{martinez2017fully} or design time \cite{singh2013mapping}.
	The latter defeats the purpose of run-time flexibility and adjustment.
	While design space exploration at runtime is infeasible due to the limited processing capacity of a supervisor, unless tight constraints are placed upon applications regarding structure and functionality \cite{martinez2017fully}.
	In practice, however, we do not have to find the singular ``best possible" solution when recovering from a fault, instead we just need a ``good enough" solutions yielded by a heuristic algorithm \cite{carvalho2007heuristics}.
	Once the system has been stabilized, ample time will be available to further optimize the thread mapping and usually this is done by the operator or flight software.

	To facilitate a heuristic approach, we first reduce these three competing objectives to a set of \emph{performance profiles}, examples of which are given in Table \ref{tab:performanceprofiles}.
	In each profile, criticality classes (essential - low) are assigned one or multiple execution modes: separate, redundant, majority voting, or with more cores, e.g., to enable Byzantine voting (referred to as NMR, TMR, DMR and separate in Table \ref{tab:performanceprofiles}).
	Duplicate assignments allow threads to be mapped in either mode, to enable mode reduction in case of resource constraints.
	For example, when running in the robustness profile, essential applications are always assigned the desired number of cores, while high-criticality applications are at least TMRed (depending on available resources).
	Other applications are preferably executed TMRed, but may be executed also DMR to retain fault detection, in case of resource exhaustion, instead of entirely de-scheduling lower criticality threads.
	Depending on mission requirements, the operator can then select the most suitable performance profile from a set of pre-generated at runtime, or could draft a new one.

	To map threads, we build a new mapping for a task using the strongest desired execution mode.
	We evaluate if this exceeds the available power budget (energy profile) or processing capacity.
	If so, we begin reducing the execution mode of tasks beginning with the last mapped and therefore lowest-criticality thread.
	If successful, we append the mapped thread to a list and proceed with the next thread.
	To minimize the amount of de-scheduled and mode reduced threads, we can sort threads of same criticality based on required processing capacity.
	Thereby, computationally expensive threads are reduced in execution mode first, freeing up larger amounts of processing resources.\looseness=-1
	
	\begin{figure}[!b]
		\vspace{-10pt}
		\centering
		\hspace{-10pt}
		\setlength{\tabcolsep}{1pt}
		\setlength{\extrarowheight}{2pt}%
		\begin{tabular}{c|c|c|c|c}
			Mode & Speed & Energy & Robustness & Function \\ 
			\hline 
			{
				\setlength{\tabcolsep}{0pt}
				\setlength{\extrarowheight}{1pt}%
				\begin{tabular}{l}
					NMR \\
					TMR \\
					DMR \\
					Separate\\
				\end{tabular}
			}
			&
			{
				\setlength{\tabcolsep}{1pt}
				\begin{tabular}{cccc}
					- & - & - & - \\
					E & H & M & L \\
					- & H & M & L \\
					- & - & - & L \\
				\end{tabular}
			}
			&
			{
				\setlength{\tabcolsep}{1pt}
				\begin{tabular}{cccc}
					- & - & - & - \\
					E & H & M & - \\
					- & - & M & L \\
					- & - & - & L \\
				\end{tabular}
			}
			&
			{
				\setlength{\tabcolsep}{1pt}
				\begin{tabular}{cccc}
					E & H & - & - \\
					- & H & M & L \\
					- & - & M & L \\
					- & - & - & - \\
				\end{tabular}
			}
			&
			{
				\setlength{\tabcolsep}{1pt}
				\begin{tabular}{cccc}
					E & H & - & - \\
					E & H & M & L \\
					- & - & M & L \\
					- & - & - & L \\
				\end{tabular}
			}
		\end{tabular}
		\caption{\emph{Performance profiles} with threads of different criticality levels (\textbf{E}ssential, \textbf{H}igh, \textbf{M}edium, \textbf{L}ow) being assigned different replication levels to enable fault detection or different voting configuration through thread replication.
		}
		\label{tab:performanceprofiles}
	\end{figure}

	If not all threads could be mapped, we can de-schedule lower-threads exceeding the compute capacity, energy constraints, or allocate less processing time to specific applications system.
	If no further mode or processing time reductions due to RT-guarantees are possible, we abort mapping, and re-traverse the list increasing execution mode, thereby undoing mode reductions due to the previous reduction steps.
	The supervisor itself only has to execute the latter part of this algorithm and perform mode and processor time reduction, or de-schedule the lowest criticality threads.
	It does not have to actually generate all these mappings as it does not enforce thread assignment in the system and only intervenes if necessary.\looseness=-1
	
	This algorithm also provides all mechanics necessary to minimize the amount of active processor cores, and as threads can be concentrated to as few tiles as possible, maximizing the number of clock-gated cores.
	Individual tasks could also signal preference for reduced processing instead of a mode reduction as the approach itself is computationally inexpensive.

	\section{Discussion and Further Applications}
	\label{sec:discussion}
	We implemented the MPSoC architecture described in Section \ref{sec:mpsoc} using Xilinx Kintex and Virtex FPGAs as well as the Zynq SDSoC platform \cite{carlson2016use}, as these are relevant for our target missions.
	However, for larger satellite platforms, this approach and architecture could very well be implemented on ASIC, and we see this as a ``big-space" variant of our approach.
	An ASIC implementation would have lower energy consumption, and allow higher clock rates due to tighter timing and shorter paths, and be less susceptible to transient faults.
	If manufactured in an inherently radiation hardened technology such as FD-SoI \cite{kochiyama2011radiation}, the system as a whole would be considerably more resistant to transient faults.
	Stage 2 would then be reduced to testing and validate tiles, while not longer being able to recover faulty tiles containing defective logic, but strong fault coverage of SEEs would be improved due to RHBM.
	
	Overall, an FPGA implementation offers stronger FDIR capabilities, better coverage for permanent faults, and high flexibility at low cost, while the ASIC variant could offer better system performance and radiation tolerance due to RHBM.
	Custom ASIC development of course is expensive and time-consuming, thus, the resulting implementation would not be a viable solution for most miniaturized satellite applications, and therefore not in the scope of this technology development project.\looseness=-1

	The relaxed cost, energy, and size constraints aboard larger spacecraft allow an implementation of our approach spanning multiple FPGAs.
	Compared to a single-chip implementation, a multi-FPGA MPSoC variant offers better scalability due to easier routing, can tolerate chip-level defects, and SEFIs to the globally shared memory controllers, these can be distributed to different FPGAs.
	Replicated thread-instances could then also be distributed across FPGAs, offering non-stop operation while one of the FPGAs undergoes full reconfiguration.
	However, our proof-of-concept is focused on a single-FPGA based prototype for nanosatellite use.
	
	Our project is focused on payload data handling and platform control for miniaturized spacecraft, and therefore application to accelerator cores supporting computational offloading is not explicitly considered in our research.
	Nonetheless, it would be very well imaginable to also protect accelerator systems using this approach, yielding at least similar benefits.
	As the structure and type of applications usually executed on accelerators is tightly constrained as compared to general purpose platform control.
	Especially synchronization for real-time applications and the impact of live-migration between tiles or state-updates on a faulty tile, become much simpler if fully deterministic application behavior is assumed, as would be the case for computational offloading.
	While our project is entirely focused on general-purpose computing instead of acceleration and computational offloading, it would be intriguing to explore this aspect further.
	
	Our existing MPSoC design utilizes an AXI interconnect, but we are currently reworking our MPSoC to instead use a NoC between tiles and shared memory controllers.
	The existing interconnect implementation allows low-latency communication, but has a large footprint, and is difficult to route\footnote{We can still achieve a functional implementation meeting timing constraints at several hundred megahertz, but the interconnect PBlock becomes disproportionately large.} for larger tile counts (without optimization, we successfully placed 8 tiles).
	A NoC instead allows not only better scalability and easier routing, but also enables the implementation of a broad variety of FT concepts such as \cite{zhou2016loft}.
	
	Tiles have direct read-only access to another tile's memory segment to allow rapid thread migration and allow real-time capacity.
	However, direct access to shared main memory is not necessary to facilitate Stages 1-3.
	The data exchange required to facilitate thread migration could very well be implemented using IPC or through sockets, when considering complex networked architectures.
	In distributed systems, our approach could thus manage threads across multiple nodes sharing data when required, at the cost of higher latency.
	
	We developed this approach to guarantee FT for opaque threaded applications on POSIX-compatible RTOS and general purpose operating systems such as RTEMS and Linux.
	However, the same functionality can also be applied to virtualized, voted systems and to runtime based platforms.
	It would be very well imaginable to implement Stage 1 within MicroPython or a hypervisor, and instead vote on Python scripts or virtual machines.\looseness=-1
	
	\section{Conclusions}
	To the best of our knowledge, the on-board computer (OBC) design presented in this paper is the first practical, non-proprietary, and affordable fault tolerance (FT) approach suitable even for very small spacecraft.
	It offers strong fault coverage, using just commercial-off-the-shelf hardware, library IP, and commodity processor cores, requiring only a single FPGA and a microcontroller based supervisor.
	The software-side FT approach outlined in Stage 1 is non-invasive to applications and the OS, therefore existing software can be reused and extended easily, while retaining real-time capabilities.
	The research presented in this paper covers the entire FDIR loop, and does not ignore or make unrealistic assumptions regarding fault detection.
	
	Our approach enables the re-use of existing development tools and IP designed for mass-produced mobile-market applications, taking an important step towards departing from the artisanal development approach in today's space computing.
	Instead of requiring new technologies to be re-invented constantly and maintained at high cost, the FT mechanics presented in this paper are flexible, which can adapt and grow with the development of computer and processor technology.
	We implemented this design on recent-generation Xilinx Virtex and Zynq/Kintex Ultrascale+ FPGAs with less than 2W power consumption (6 tiles on Kintex XCKU5P) and validated the approach through fault injection.
	
	We do not just enable FT for a satellite class which today is considered unreliable, but also enhance the fault coverage capabilities of OBCs in larger spacecraft, and other applications with similar constraints and fault profile.
	Our approach facilitates majority voting through dynamic, replicated thread groups mapped to the available processor cores dynamically at runtime, instead of hardwiring them.
	Thus, all processing capacity, including spares, are part of a shared resource pool.
	Therefore, spare resources can be used more efficiently, and allowing idle compute capacity to be used productively until it is needed for fault coverage.
	An OBC running the presented hybrid hardware-software FT approach can adapt to varying mission requirements regarding adjusting the OBC transparently at run-time, trading processing capacity for reduced energy consumption or increased fault coverage.
	
	\scriptsize
	\section{Acknowledgements}
	This approach was developed for a 4-year European Space Agency (ESA) NPI project, and we are implementing a prototype jointly with two industrial partners.
	We would like to thank Gianluca Furano, Giorgio Magistrati, Antonios Tavoularis and Kostas Marinis at ESTEC/TEC-EDD for their support and feedback. 
	We thank ARM Ltd. and Softbank for making available the relevant processor and infrastructure IP. 
	N.M. Murillo acknowledges funding through the European Union A-ERC grant 291141 CHEMPLAN, by the Netherlands Research School for Astronomy (NOVA), and a Royal Netherlands Academy of Arts and Sciences (KNAW) professor prize.

	\bibliographystyle{IEEEtran}
	\bibliography{ahs2018}
	
\end{document}